\documentclass[11pt,a4paper]{article}
\usepackage[left=1in, right=1in, top=1in, bottom=1in]{geometry}
\usepackage{graphicx}
\usepackage{bm}
\usepackage{amsmath}
\usepackage{amssymb}
\usepackage{bbm}
\usepackage{mathabx}
\usepackage{caption}
\usepackage{subcaption}
\usepackage{booktabs} 
\usepackage{float}
\usepackage[final]{pdfpages}
\usepackage{listings}
\usepackage{subfiles}
\usepackage[mode=buildnew]{standalone}
\usepackage{tikz}
\usetikzlibrary{calc,arrows,patterns,intersections}
\usepackage{pgfplots}
\usepackage{xcolor}
\usepgfplotslibrary{fillbetween}
\usepackage[colorinlistoftodos,prependcaption,textsize=tiny]{todonotes}
\usepackage{shellesc}
\usepackage{authblk}
\usepackage[symbol]{footmisc}

\renewcommand{\thefootnote}{\fnsymbol{footnote}}

\newcommand{\astfootnote}[1]{%
    \let\oldthefootnote=\thefootnote%
    \setcounter{footnote}{0}%
    \renewcommand{\thefootnote}{\fnsymbol{footnote}}%
    \footnote{#1}%
    \let\thefootnote=\oldthefootnote%
}

\usepackage[backend=biber,
            style=numeric, 
            sorting=none,
            abbreviate=false,
            dateabbrev=false,
            alldates=long]{biblatex}
\renewbibmacro*{volume+number+eid}{%
    \printfield{volume}%
    \setunit*{\addnbspace}
    \printfield{number}%
    \setunit{\addcomma\space}%
    \printfield{eid}
}
\DeclareFieldFormat[article]{number}{\mkbibparens{#1}}

\renewbibmacro*{in:}{%
    {\printtext{\intitlepunct}}
}

\usepackage{color}   
\usepackage{hyperref}
\hypersetup{
    colorlinks=true, 
    linktoc=all,     
    linkcolor=blue,  
}

\begin{document}
\title{
Ecosystem for Demand-side Flexibility Revisited: \\ The Danish Solution
}
\author[1,2]{Peter A.V. Gade\footnote{Corresponding author. Tel.: +45 24263865. \\ Email addresses: pega@dtu.dk (P.A.V. Gade), Trygve.Skjotskift@ibm.com (T. Skjøtskift), hwbi@dtu.dk (H.W. Bindner), jalal@dtu.dk (J. Kazempour).}}
\author[2]{Trygve Skjøtskift}
\author[1]{Henrik W. Bindner}
\author[1]{Jalal Kazempour}
\affil[1]{Department of Wind and Energy Systems, Technical University of Denmark, Kgs. Lyngby, Denmark}
\affil[2]{IBM Client Innovation Center, Copenhagen, Denmark}
\renewcommand\Affilfont{\itshape\small}

\date{\today}
{\let\newpage\relax\maketitle}

\section*{Abstract}


Denmark has recently set a legislation called \textit{Market Model 3.0}  \textcolor{black}{to make the ecosystem for demand-side flexibility more attractive to stakeholders involved.} The main change is to relax the previous mandate that required each aggregator to be associated with a retailer and a balance responsible party. We explain the rationale behind such a change and its implications, particularly on  the pre-qualification of  \textcolor{black}{demand-side} portfolios providing ancillary services. 


\section{Introduction}

\begin{figure}[H]
    \centering
    \includegraphics[width=0.5\textwidth]{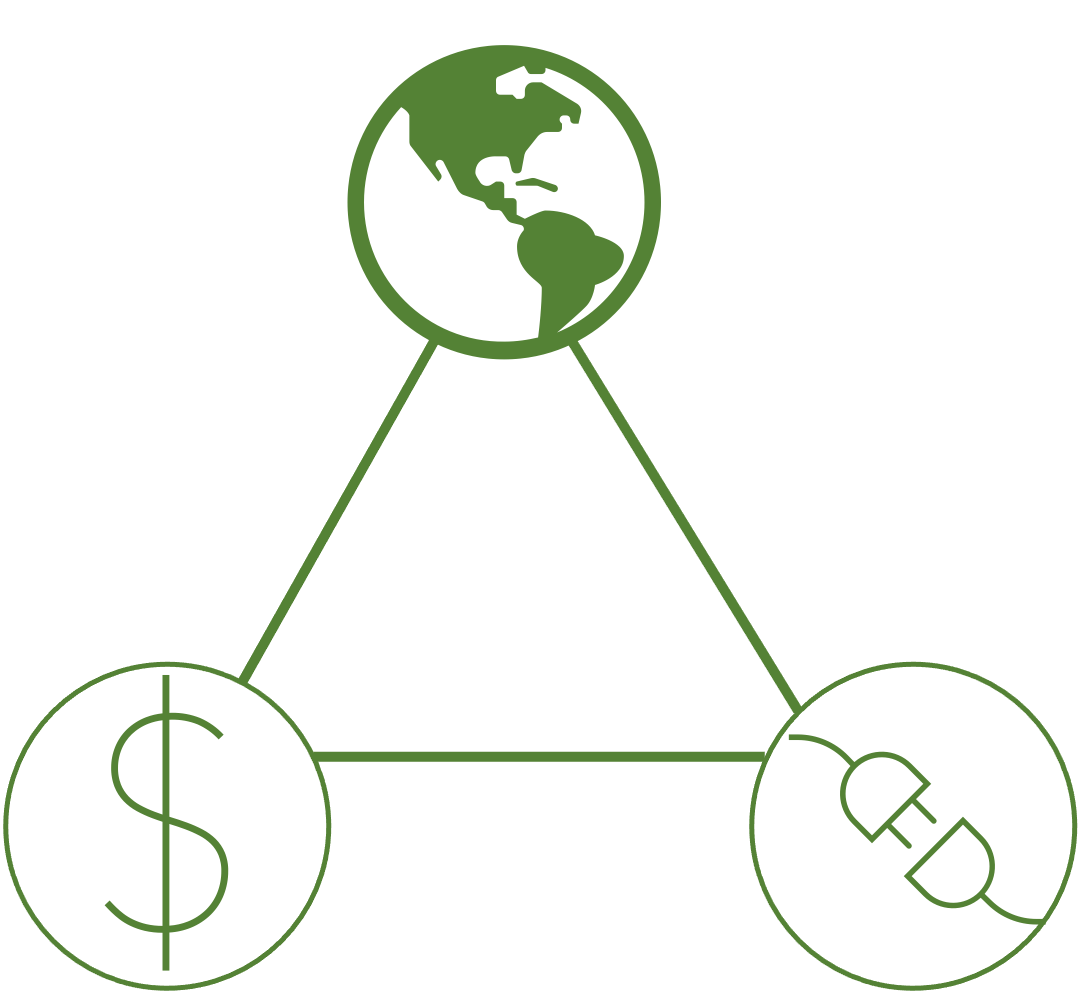}
    \caption{The energy trilemma: Energy should be affordable, green, and with a high security of supply.}
    \label{fig:energy_trilemma}
\end{figure}

There is a global effort towards reducing carbon dioxide emissions. In Denmark, the goal is to reduce emissions by 70\% in 2030 compared to the 1990 level. It is widely believed that renewable-based electrification of transportation and heating sectors will play a central role for reaching such a target. In particular, 1.5 million cars as well as 210,000 heat pumps in Denmark must be electric by 2030 \cite{Ostergaard2021}. However, such a massive electrification will make the power grid  further stressed, potentially causing a broad range of technical challenges such as line congestion and nodal voltage issues. In addition, the increased level of power supply uncertainty due to growing integration of stochastic renewable power sources may compromise the supply security. All these challenges need to be properly resolved while keeping the price of electricity at a reasonable level. This constitutes the energy trilemma as shown in Figure \ref{fig:energy_trilemma}, implying that the electricity must be affordable, with little to ideally no carbon emissions, and with a high security of supply.



The key to integrate large-scale stochastic renewable energy sources and to massively electrify transportation and heating sectors is the availability of sufficient resources of operational \textit{flexibility} in the power grid. Flexibility is needed at all levels and in all parts of the network. Therefore, all potential resources of flexibility must be made available. Flexible resources enable the power grid operator in both transmission and distribution levels to adjust any imbalances caused by wrong forecasts and fluctuations of renewable power generation or demand, and to safely operate the power grid by relieving line congestion and voltage issues. The flexibility in the supply side is inherently limited, and it becomes even  scarcer by phasing out conventional generators. Meanwhile, large-scale storage technologies are still immature or costly. This leaves us with the need for harnessing flexibility mainly from Distributed Energy Resources (DERs) including flexible loads. 
Flexible power consumption is often referred to as \textit{demand-side flexibility}, which implies the capability of electric power demands to directly or indirectly adjust their consumption level according to an external signal provided by the power grid operator. 

Demand-side flexibility has shown great promise in many studies and is widely believed to be a vital part of the solution to the energy trilemma. The world energy outlook estimates a need for 500 GW capacity of demand-side flexibility in 2030 \cite{WorldEnergyOutlookDemandResponse}.
There are several reasons that the demand-side flexibility is becoming an appealing option. Among others, one main reason is that Internet of Things (IoT) infrastructure has become cheaper over the years which makes it worthwhile communicating with multiple DERs in a coordinated manner.
Despite all advantages of DERs, they have not been widely adopted yet on a grand scale, due to legislation and technical barriers. The legislation barrier is that the current regulations have traditionally been centered around conventional fossil-fuel energy generation and the development of electricity markets. 
From the technical perspective, the main barrier is that the utilization of demand-side flexibility is challenging to assert and verify, owing to the stochastic nature of flexible loads. 


The demand-side flexibility aggregator, in short \textit{aggregator}\footnote{Sometimes referred to as a flexibility service provider.}, is in charge of aggregating the flexibility of DERs and shaping the aggregated flexibility in the form of various ancillary services. Aggregators have potential to profoundly impact the ecosystem of participants involved in the delivery of electricity and responsibility of imbalances in electricity production and demand. However, the lack of revised regulations has prevented demand-side flexibility and aggregators from operating with a significant importance to the power grid. The Danish regulatory entity has realized that changes are necessary to the policies and legislation that govern how the current market framework is set with respect to the balancing of the power grid. 
It has called for a change in regulations and policies that explicitly cater for demand-side flexibility from the bottom up.

In a pioneering effort, Denmark has now become the first country in the world to officially address these issues through an initiative called \textit{Market Model 3.0} \cite{energistyrelse:Marketmodel3.0}. It was proposed and outlined in 2021 by Energistyrelsen (Danish Energy Agency) from which the Danish regulating entity, Forsyningstilsynet, revised the regulation and policies after consulting with the Danish transmission system operator, Energinet, who is responsible for the implementation. Market Model 3.0 was drafted using all the learnings and experiences made in various research and demonstration projects, funded by Danish and European grant agencies, related to smart grids, demand response, and power flexibility. 
Market Model 3.0 is a first-of-its-kind ecosystem that rethinks the electricity market design with a specific focus on how demand-side flexibility can contribute to the future power system. It contains tangible solutions to some of the barriers hindering aggregators to unlock and utilize the demand-side flexibility. 

The rest of the paper is organized as follows: Section 2 describes the status quo of the ecosystem of participants in the power system before Market Model 3.0, especially with a focus on the role of aggregators. This section also explains the main grid services that demand-side flexibility can provide. Section 3 presents Market Model 3.0, and explains how it changes the role of aggregators in the existing ecosystem of market participants, and how it incentivizes flexibility provision from potential aggregators and new participants. Lastly, Section 4 discusses the implementation of Market Model 3.0 and its implications.

\section{Ecosystem of flexibility provision before Market Model 3.0}
In current power systems, flexibility is being provided to the Transmission System Operator (TSO), e.g., Energinet in Denmark, in the form of various \textit{ancillary services}. These services are different in terms of how fast they can respond, if activated. Separate markets have been designed for pricing different ancillary services, where the TSO remunerates the flexibility providers for their services. In this section, we discuss the current ecosystem and challenges with the current approach of providing demand-side flexibility, which serves as a motivation for outlining the initiative of Market Model 3.0. Note that we do not discuss on potential flexibility services needed for distribution systems. 


\subsection{Status Quo in Denmark}

The Nord Pool day-ahead electricity market works like an auction of sellers and buyers with a settlement according to the merit-order principle. This market is cleared every day where producers offer energy to sell and retailers, and in general Balance Responsible Parties (BRPs), buy electricity for their consumers. In addition, there is a continuous intra-day market where participants can adjust their day-ahead schedule with direct bids and offers cleared with a pay-as-bid pricing scheme. In the real-time operation, there could be a mismatch between production and demand, which is the responsibility of the TSO to ensure there is a balance between electricity production and demand. The TSO is also in charge of the operation of the transmission grid, which could be a challenging task as day-ahead and intra-day market outcomes are not necessarily feasible due to using a simplified representation of the transmission network. 
The TSO traditionally operates the system by buying power reserve capacity which can be activated, if needed. For example, conventional generators might reserve some capacity for providing ancillary services to the TSO by up- or down-regulating their production, as will be explained later.

\begin{figure}[H]
    \centering
    \includegraphics[width=1.1\textwidth]{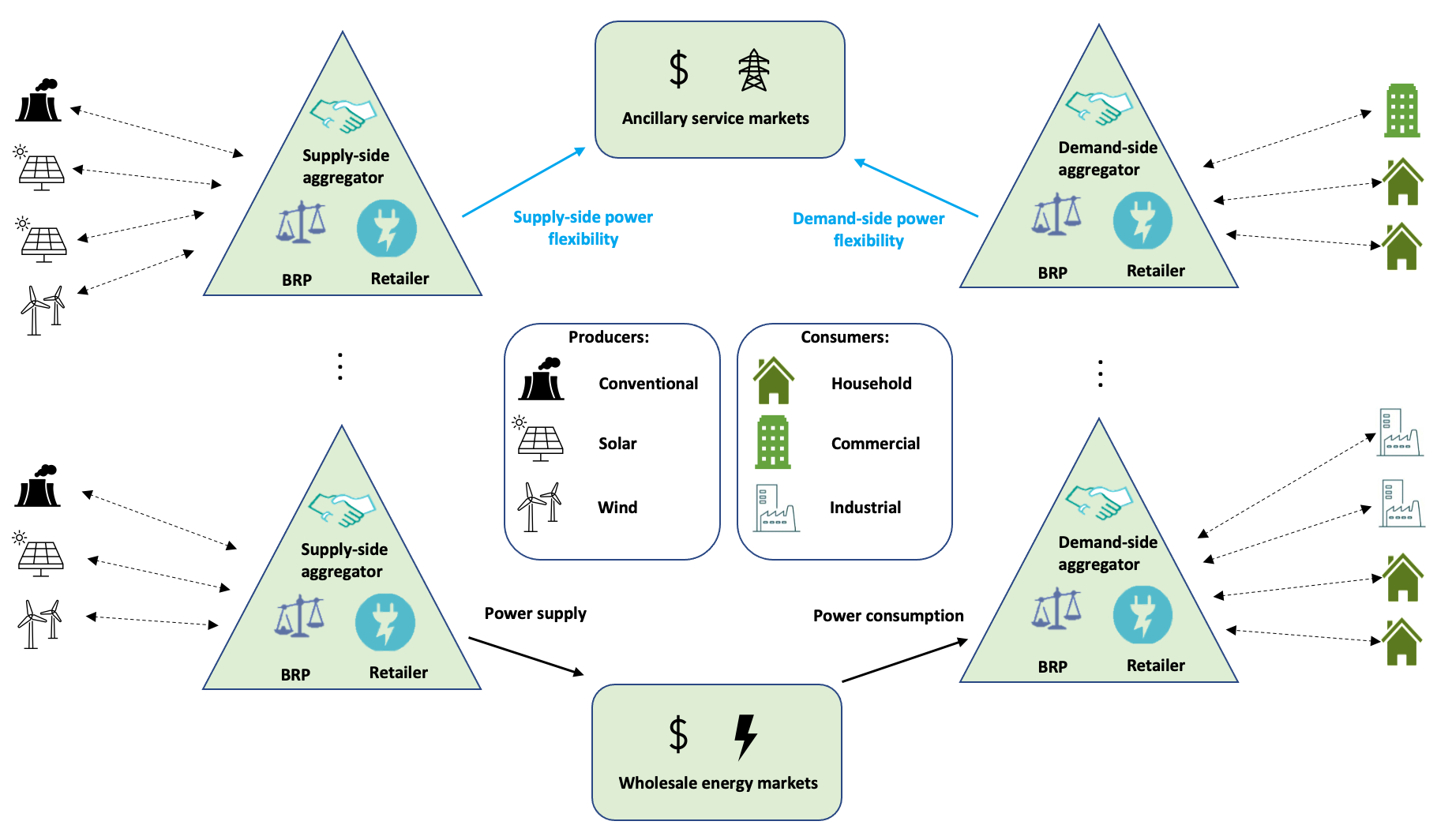}
    \caption{Ecosystem of energy and flexibility provision in the Danish power system \textit{before} Market Model 3.0. Here, an aggregator (either in supply- or demand-side) is associated with a BRP and a retailer (see triangles). Market Model 3.0 is going to relax the need for such an association. Recall that the focus of this paper is the demand-side flexibility.}
    \label{fig:aggregator_role_today}
\end{figure}

Figure \ref{fig:aggregator_role_today} readily shows the Danish power system ecosystem and the role of different entities. Producers, either renewable or conventional, sell electricity in the wholesale energy markets (i.e., day-ahead and intra-day). Consumers, either residential or commercial or industrial, have a retailer who buys electricity for them from the producers as indicated by the black arrows. Each retailer is associated with a BRP that has the financial responsibility for any deviations to the electricity bought by the retailer and the actual demand or actual power production of the retailer's consumers or producers. The electricity for the consumers is bought in the wholesale energy markets by the BRP on behalf of the retailers.\footnote{In this context, the retailer is simply the administrator of the end consumers with no balance responsibility. In the US, a utility or a retail electricity provider is the entity that buys electricity and is responsible for potential imbalances.}  \textcolor{black}{It is worth mentioning that each aggregator-retailer-BRP triplet has a unique set of consumers which does not intersect with other consumer sets.} This prevents imbalances caused by other participants. In the context of demand-side flexibility, an aggregator has to fit into this ecosystem which requires a strong business-to-business relationship with BRPs and/or retailers. The aggregator is therefore implicitly defined by the ecosystem. The demand-side flexibility can then be sold as ancillary services to the TSO as indicated by the blue arrows.

\subsection{Ancillary service provision}

Ancillary services for the TSO can be divided into three categories: frequency services, voltage services, and black-start services --- see \cite{kirby2007ancillary} for a thorough description. Frequency services stabilize the power grid such that electricity production equals demand when the grid frequency constant at the nominal frequency\footnote{The nominal frequency is 50 Hz in Europe and 60 Hz in the US.} which corresponds to equal power production and consumption in a synchronous area. There is a constant need to balance the power grid at all times, and frequency service provision helps mitigate intermittency in the electricity supply and changes in demand. Voltage services are used to adjust voltage levels in the transmission grid by deploying reactive power. Black-start services help restart the whole system after a blackout. Our focus in this paper is on frequency services. 

The procurement process of frequency services differs depending on the type of reserve. Frequency services mainly consist of three types of reserves: primary reserves, secondary reserves, and tertiary reserves; the main difference being the time scale they operate in. In Denmark and generally in Europe, they correspond to Frequency Containment Reserve (FCR), Automatic Frequency Restoration Reserve (aFRR), and Manual Frequency Restoration Reserve (mFRR). Both FCR and aFRR are automatic and respond to the frequency in the power grid directly. The FCR is deployed before aFRR, usually within a few seconds, whereas aFRR is deployed after 30 s. The mFRR is manually deployed either reactively or proactively. Reactive deployment refers to the case mFRR is activated after FCR and aFRR, whereas proactive deployment is in anticipation of imbalances as per the TSO's discretion. In Denmark, mFRR is primarily deployed in a proactive manner \cite{energinet:Systemydelser}. 

\begin{figure}[t]
    \centering
    \includegraphics[width=0.9\textwidth]{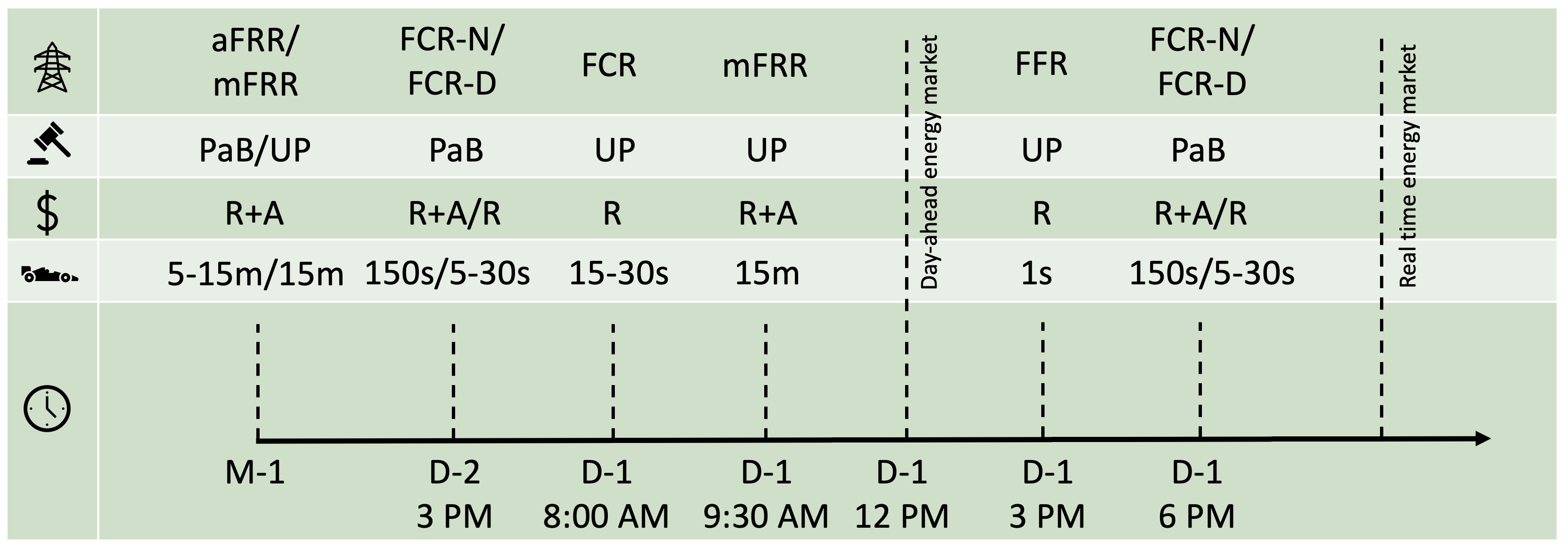}
    \caption{Timeline for reservation auctions for all ancillary services in Denmark, inspired by \cite{energinet:GennemgangAfMarkeder}. Service mFRR has monthly and daily auctions. Both FCR-N and FCR-D have two auctions, running two days and one day before delivery. The auction scheme is shown together with the origin of payment (third row) and the speed of full response (fourth row). For aFRR, the speed of full response is 15 min in DK1 and 5 minutes in DK2, and will be 5 min in DK1 from 2024. \textit{Abbreviations:} PaB: pay-as-bid. UP: uniform pricing. R: reservation payment. A: activation payment. M: month. D: day. m: minutes. s: seconds. }
    \label{fig:as_overview_custom}
\end{figure}

Figure \ref{fig:as_overview_custom} shows the timeline for the procurement of ancillary services in Denmark. Denmark has two synchronous areas, DK1 (western and central Denmark) and DK2 (eastern Denmark). DK1 is connected to continental Europe which is a very robust system from a frequency stability perspective, and therefore FCR, aFRR, and mFRR services are sufficient. DK2, however, is connected to the Nordic area which is a smaller system compared to the continental Europe, and therefore additional frequency-supporting ancillary service markets, including FCR for normal operation (FCR-N), FCR for disturbances (FCR-D), and Fast Frequency Reserve (FFR), have been designed to stabilize the frequency. As mentioned earlier, this paper focuses on FCR, aFRR, and mFRR services only, although Figure \ref{fig:as_overview_custom} presents all ancillary services in DK1 and DK2 for completeness. 

According to Figure \ref{fig:as_overview_custom}, some ancillary services have multiple auctions, i.e., monthly and daily or two daily auctions. Markets with a non-negligible energy delivery have both reservation (per MW) and activation (per MWh) payments. All markets consider either a pay-as-bid or a uniform pricing scheme. The real-time activation payment is obtained by uniform pricing. Note that mFRR services are being procured by the Danish TSO before the day-ahead market clearing. This is challenging since the TSO does not know the production and demand schedules in advance. The procurement process for each market is an important factor for an aggregator when delivering flexibility. For example, if day-ahead prices are not known before bidding mFRR capacity, it might be necessary to account for this uncertainty.

\begin{figure}[t]
    \centering
    \includegraphics[width=0.9\textwidth]{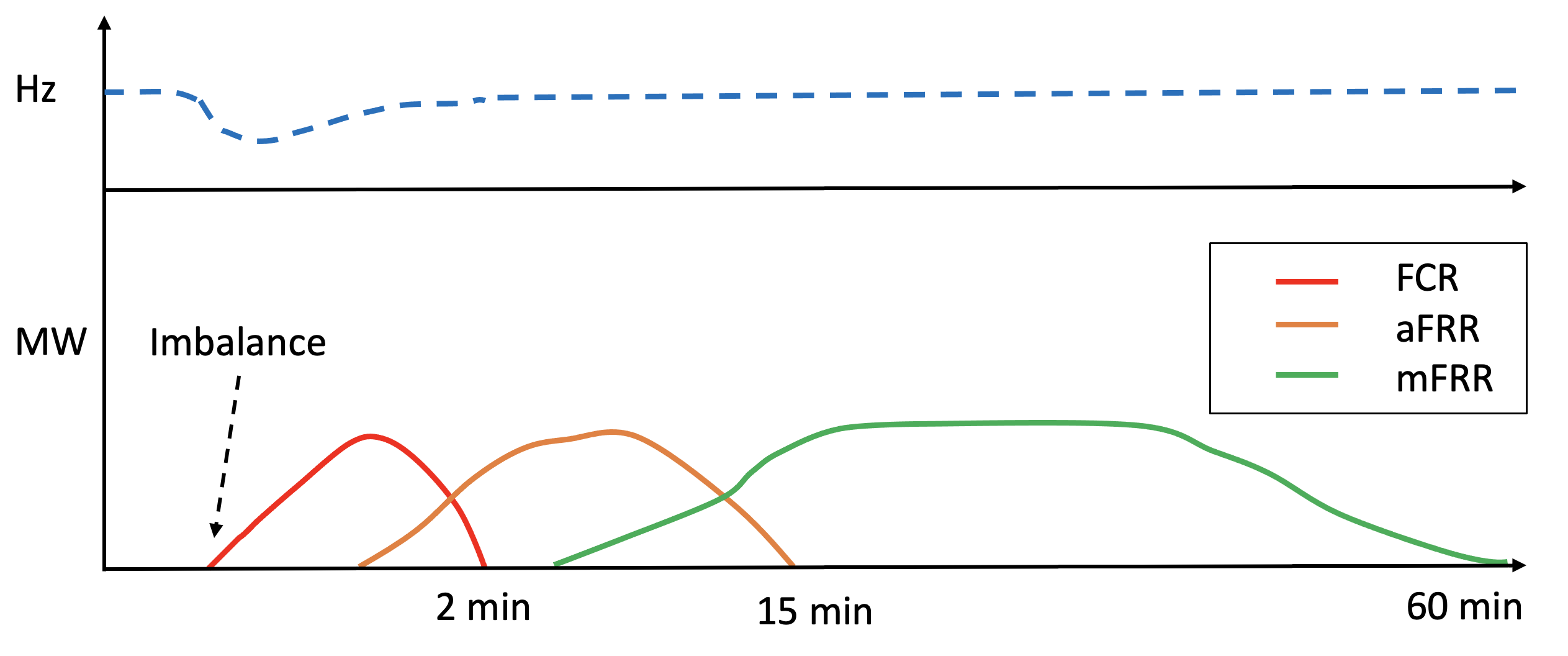}
    \caption{Timeline for the activation of FCR, aFRR, and mFRR in Denmark \textcolor{black}{(lower plot)}, following an imbalance that causes a frequency drop below the nominal frequency \textcolor{black}{(upper plot)}. This is an example of reactive deployment of reserves.}
    \label{fig:reserve_market_timelines}
\end{figure}

Figure \ref{fig:reserve_market_timelines} illustrates the timeline for the activation process of three different reserves: when an imbalance happens, the frequency deviates. For example, in the case of the lack of sufficient production to supply demand, the frequency drops below the nominal frequency, and therefore generators and demands with reserve capacity deliver up-regulation services. Generators deliver up-regulation by increasing their power production level, whereas demands do so by decreasing their consumption level. First, FCR is provided from reserve providers that can respond within seconds to frequency deviations. Then, other providers respond more slowly by providing aFRR. Lastly, mFRR is deployed for about an hour to keep the frequency stable.

In Denmark today, the TSO buys 83 MW capacity for FCR, 100 MW for aFRR, and about 1 GW of mFRR \textcolor{black}{combined, for all auctions} \cite{energinet:Systemydelser}. In total, this corresponds to about 20\% of the Danish power consumption during peak hours. Traditionally, BRPs deliver these reserves on behalf of the owners of power production plants or demands. They are paid according to the capacity they make available. For example, a hydro generation plant might have a nominal capacity of 5 MW, but only sells 4 MW in the day-ahead market while keeping 1 MW in reserve for mFRR. Some reserves, such as aFRR and mFRR, have a non-negligible delivery of energy when activated. These types of reserves therefore require a BRP for settlement purposes. As such, BRPs are paid according to available capacity and also for activation of that capacity when the energy delivered is non-negligible.

\begin{figure}[t]
    \centering
    \begin{subfigure}[t]{0.49\textwidth}
        \centering
        \begin{tikzpicture}[scale=1.0]
    \begin{axis}[
        ylabel=$\textrm{Power}$ (kW),
        ymin=5,ymax=20,
        xmin=0, xmax=6.5,
        grid=both,
        xticklabels={,,12:00,,,,12:01,},
        yticklabels={,,,},
        width = \axisdefaultwidth,
        domain=1:5
      ]
      \addplot[name path=Baseline,smooth,dashed,black, very thick] plot coordinates {
          (1,10)
          (5,10)
        };
      \addlegendentry{Baseline}
      \addplot[name path=realized, smooth, red, mark=x, mark size=2pt] plot coordinates {
          (1,10)
          (1.5,11.5)
          (2,12)
          (2.4,8)
          (3,11.3)
          (3.6, 12)
          (4,12.2)
          (4.4,8.8)
          (5,11.0)
        };
      \addlegendentry{Realized}

      \addplot[blue!50, opacity=0.2] fill between[of=Baseline and realized];
      \addlegendentry{Energy}

      \addplot[only marks, mark=x, orange, mark size=4.0pt] coordinates {(2.2,10) (2.77,10) (4.25,10) (4.8,10)};

    \end{axis}
  \end{tikzpicture}
        \caption{FCR activation: Realized power output varies around the baseline schedule according to the frequency deviation. When the frequency is below 50 Hz, there is a need for up-regulation which means the generator produces more power, and vice-versa for down-regulation. When the frequency is 50 Hz (orange crosses), the power output should equal the baseline. A one-minute snapshot of the whole reservation hour is shown.}
        \label{fig:FCR_conventional_generator_v2}
    \end{subfigure}
    \hfill
    \begin{subfigure}[t]{0.49\textwidth}
        \centering
        \begin{tikzpicture}[scale=1.0]
    \begin{axis}[
        ylabel=$\textrm{Power}$ (kW),
        ymin=0,ymax=20,
        xmin=0, xmax=6.5,
        grid=both,
        xticklabels={,,12:00,,,,13:00,},
        ytick={2.5, 5, 10, 15, 20},
        yticklabels={$ \textrm{Baseline} - \Delta \textrm{P}^{\downarrow}$,$ \textrm{Baseline} - \frac{2}{3}\Delta \textrm{P}^{\downarrow}$,, , },
        width = \axisdefaultwidth,   
        domain=1:5  
     ]
    \addplot[name path=Baseline,smooth,very thick,dashed,black] plot coordinates {
        (1,10)
        (5,10)
    };
    \addlegendentry{Baseline}

    \addplot[name path=realized, color=red, mark size=1.5pt]
        plot coordinates {
            (1,5)
            (2,5)
            (2,5)
            (2,10)
            (4,10)
            (4.0, 2.5) 
            (4,2.5)
            (5,2.5)
        };
    \addlegendentry{Realized}
    \addplot[blue!50, opacity=0.2] fill between[of=Baseline and realized];
    \addlegendentry{Energy}

    \end{axis}
 \end{tikzpicture}
        \caption{mFRR activation (here down-regulation): The mFRR capacity of the generator is $\Delta\textrm{P}^{\downarrow}$. At 12:00, two-thirds of the reserve is activated and deployed for 15 min. From 12:15-12:45, there is no reserve activation and therefore the generation level follows the schedule. From 12:45-13:00, the entire reserve capacity is activated.}
        \label{fig:mFRR_conventional_generator_downregulation_v2}
    \end{subfigure}
    \caption{A sample activation of a typical conventional generator for FCR (left plot) and mFRR (right plot) services with reservation between hours 12:00 and 13:00. Baseline: total schedule of the generator in the day-ahead and intra-day markets. Realized: realized power output of the generator after reserve activation. Energy: energy delivered for reserve provision.}
    \label{fig:generator_flexibility}
\end{figure}
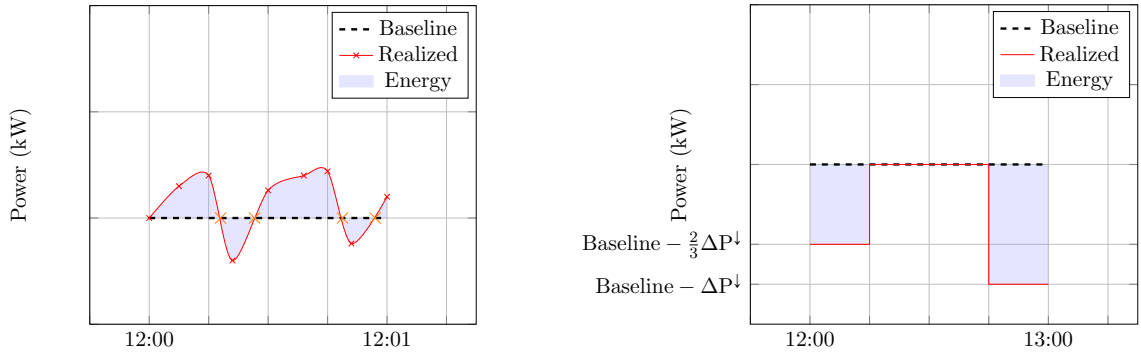

The provision of ancillary services must be \textit{pre-qualified} by the TSO, which means that a large degree of feedback and control is required of the generator or demand portfolio that is to deliver ancillary services. The pre-qualification is effectively a demonstration that consists of a series of tests used to verify that ancillary services can be provided as prescribed in the service specification \cite{energinet:prequalification}. 
Ancillary services have mainly been procured from few conventional generators which are able to provide flexible capacity to the energy system \cite{kirby2007ancillary}. Figure \ref{fig:generator_flexibility} shows conceptually how a typical conventional generator provides flexibility in form of FCR and mFRR. This figure does not consider aFRR since its provision operates on a time scale slower than FCR, but faster than mFRR. Hence, FCR and mFRR represent the two edge cases.

Figure \ref{fig:FCR_conventional_generator_v2} illustrates the case of FCR provision where the generator's production schedule from trading in the day-ahead and intra-day markets constitutes the \textit{reference baseline} from which the flexibility is provided: when the generator produces more than the baseline, it is because the frequency in the power system is below 50 Hz, and vice versa when it produces less than the reference. 
Figure \ref{fig:mFRR_conventional_generator_downregulation_v2} depicts the case of mFRR provision (down-regulation). For example, in the first 15-minute activation period the TSO only requires two-thirds of the capacity while it requires the full capacity in the last activation period. Usually, mFRR is activated for a duration of one hour but with the recent introduction of the Nordic Balancing Model, mFRR will be activated in 15-minute time blocks \cite{NordicBalancingModel, NordicBalancingMarkets}.

\subsection{Challenges with conventional ancillary service provision}

The crucial aspects of ancillary service provision from conventional generators are twofold: 1) setting the reference from which the ancillary service response is delivered from, and 2) verifying service provision and economic settlement which should be carried out ex-post. The ancillary service provision depicted in Figure \ref{fig:generator_flexibility} requires that the reference (i.e., the baseline schedule) is \textit{known}. In the case of aggregators, many studies in the literature consider a similar assumption on having a known baseline for production and consumption levels \cite{morales2013integrating, petersen2013market}.

The traditional ecosystem (where the baseline is assumed to be known) constrains flexibility aggregators, since they must know the energy bought for their demand-side portfolio. The barrier to entry for new aggregators is therefore high as it requires a business relationship to existing retailers and/or BRPs. \textcolor{black}{Hence, a new ecosystem is needed where aggregators are not assumed to know the reference power schedule of their demand-side portfolio}. This is exactly one of the initiatives in Market Model 3.0 as will be explained in the next section.



\section{Initiative in Denmark: Market Model 3.0}

Market Model 3.0 realizes past decades of research in demand-side flexibility by explicitly outlining a market framework designed to accommodate demand-side flexibility in the energy system. It represents the first steps towards an integrated smart grid to support the green transition. Some countries have already established demand response markets. For example, TenneT (TSO in the Netherlands) has designed a passive balancing mechanism where BRPs are paid if they provide flexibility upon a broadcasted signal from the TSO \cite{TENNETNL}. In California, a new initiative called \textit{Market Access} tries to fast-track solutions for demand response due to past years summer blackouts by reducing the complexity of utility programs \cite{California}. Nonetheless, none of these initiatives directly accommodates or substantiates demand-side flexibility from a holistic perspective that is integrated into the existing market framework.

As explained, the ecosystem of current participants and actors in the energy system does not necessarily allow or incentivize demand-side flexibility for new actors wishing to create new business models built upon demand-side flexibility \cite{biegel2014value}. This is due to both legislation and technical issues. From a technological point of view, it is difficult to provide demand-side flexibility for ancillary services, especially for FCR at fast time scales due to the need for coordinated control of multiple DERs. In what was traditionally a huge barrier for demand-side flexibility, it has now become much cheaper to wirelessly communicate with assets through IoT and the marginal cost of onboarding an additional asset or KW has decreased. A significant amount of research and efforts have gone into demand-side flexibility, however those efforts will be fruitless unless the ecosystem fundamentally changes with respect to legislation. The main four barriers in this regard are listed below:

\begin{itemize}
    \item Current pre-qualification requirements for ancillary services are very strict and particular in terms of the required response. They are originally made for conventional generators. This includes strict minimum requirements of power measurement resolution and reporting based on existing Supervisory Control and Data Acquisition (SCADA) infrastructure which is difficult to adhere  for flexible loads.
    \item There is a need for a BRP for ancillary services with non-negligible energy delivery upon activation. This greatly reduces the attractiveness of providing flexibility since a business-to-business relationship must first be formed. In Denmark, there are approximately fifty BRPs \cite{energinet:BRP_list}.
    \item The strong dependence of aggregators on the existing ecosystem makes it less profitable to deliver ancillary services since the profit has to be divided between several entities.
    \item There is a lack of market structure in the distribution grid, causing an uncertainty in how demand-side ancillary service provision will be compensated, and how it should be delivered.
\end{itemize}

Market Model 3.0 addresses these issues with the following three initiatives: 
\begin{itemize}
\item Re-defining the aggregator's role by easing access to ancillary service markets,
\item Changes to pre-qualification of demand portfolios delivering ancillary services,
\item Changes to the legislation of Distribution System Operator (DSO) monopolies to incentivize data transparency and revealing the state of the distribution grid. 
\end{itemize}

In this work, the focus will only be on the first two issues adopting a transmission perspective due to the lack of \textcolor{black}{local flexibility markets} in the distribution grid. \textcolor{black}{There is a need for local flexibility in the distribution grid as well \cite{jin2020local, bouloumpasis2019congestion}.}

In the following, we first describe the ecosystem in Market Model 3.0 with a comparison to the previous ecosystem. We explain how the aggregator's role is \textit{explicitly} defined by Market Model 3.0 as opposed to today. Finally, we illustrate how the demand-side flexibility provision will work in Market Model 3.0 for FCR and mFRR.

\subsection{Ecosystem of flexibility provision in Market Model 3.0}
\begin{figure}[t]
    \centering
    \includegraphics[width=0.99\textwidth]{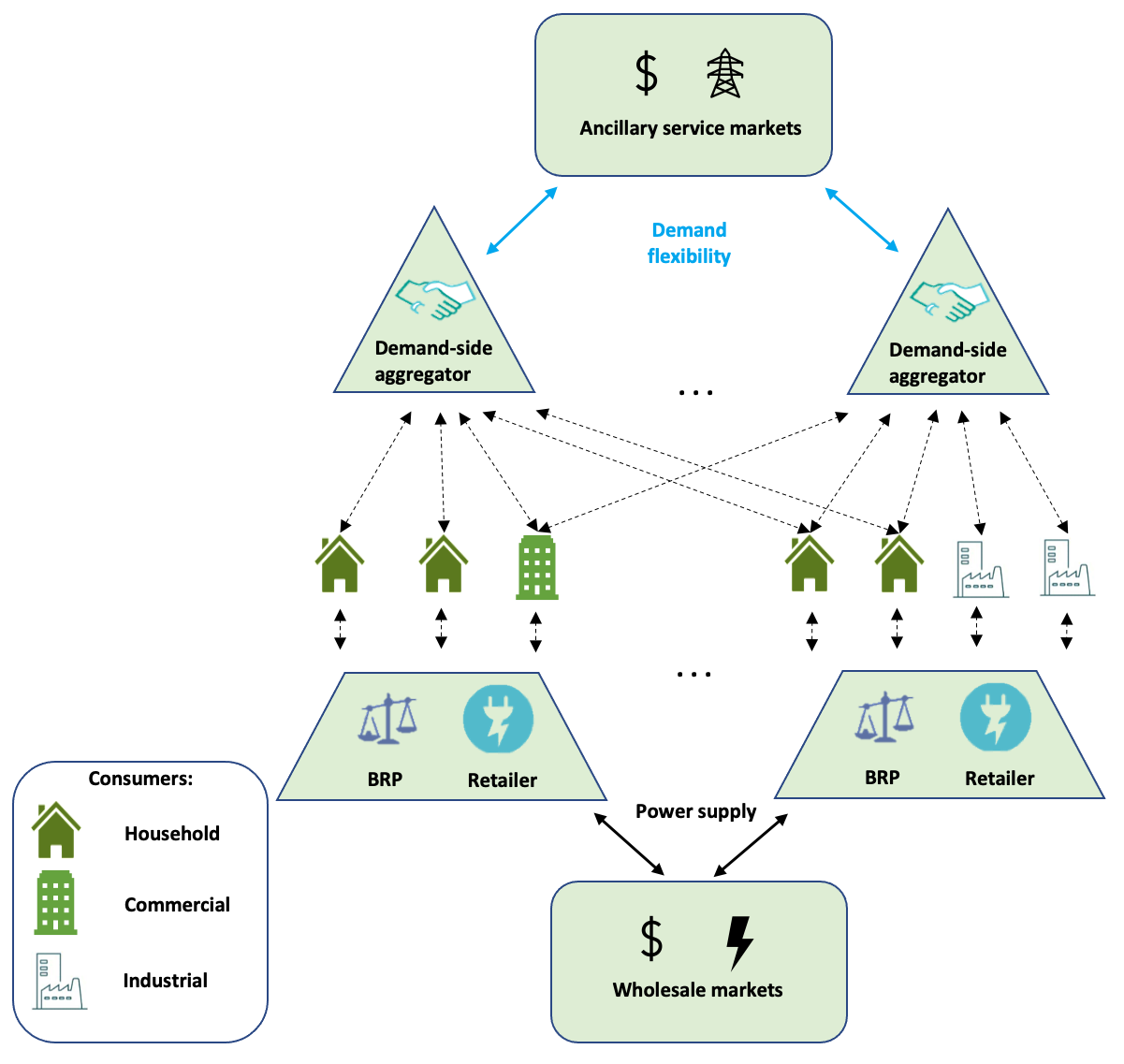}
    \caption{Ecosystem of flexibility provision in Market Model 3.0 from the demand-side perspective. \textcolor{black}{When compared to the ecosystem before Market Model 3.0 in Figure \ref{fig:aggregator_role_today}, the demand-side aggregator has now become independent and is allowed to have a demand-side portfolio across multiple retailer-BRP doublets.}}
    \label{fig:aggregator_role_marketmodel}
\end{figure}

Figure \ref{fig:aggregator_role_marketmodel} illustrates how the aggregator's role changes explicitly with Market Model 3.0 compared to status quo as previously shown in Figure \ref{fig:aggregator_role_today}. An aggregator is independent of incumbent BRPs and retailers, and interacts with ancillary service markets directly. Thus, an aggregator can have a demand portfolio where the consumers can have multiple underlying BRPs that are financially responsible for imbalances of their respective demands. In this way, aggregators does not legally need to go through BRPs to provide ancillary services with non-negligible energy delivery. However, as described later, this setup must be fair towards any imbalances incurred to other BRPs by an aggregator.

For example, Market Model 3.0 allows a manufacturer of smart refrigerators to become an aggregator by utilizing the flexibility in the refrigerators for ancillary services, perhaps as part of a service agreement when selling the refrigerator in the first place. In this case, the manufacturer aggregates flexibility from multiple consumers who will inevitably have different retailers, and thereby BRPs, but the aggregator only has to pass the pre-qualification towards the TSO, not including the consumer's BRP in the bidding, reservation, and activation processes. 

\subsection{Flexibility provision in Market Model 3.0}

Given that the flexibility aggregator of a demand portfolio is assumed to be independent of the BRP and the retailer as indicated in Market Model 3.0, the reference power from trading in the day-ahead and intra-day markets is generally \textit{unknown} as opposed to the case of a conventional generator as previously illustrated in Figure \ref{fig:generator_flexibility}. We will show later that a different baseline must be computed to assert that the required response has been carried out. In the following, similar to the case of a conventional generator, we only consider FCR and mFRR services with respect to the time scale they are operated in.


\subsubsection{FCR and mFRR provision}

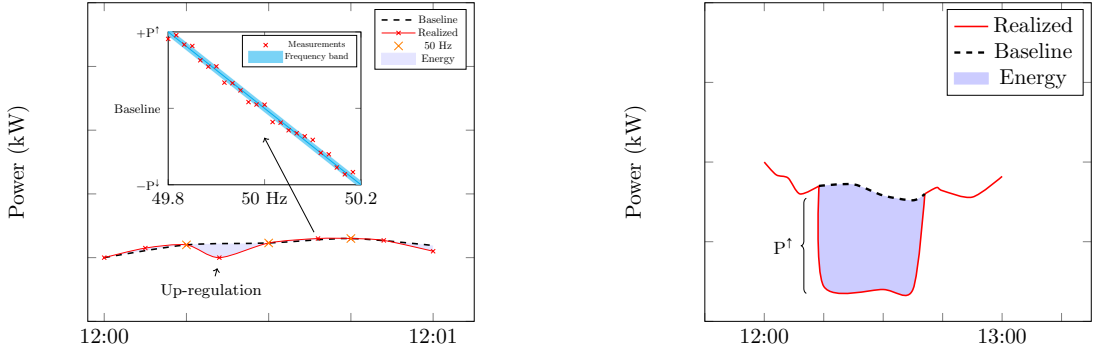
\begin{figure}[t]
    \centering
    \begin{subfigure}[t]{0.49\textwidth}
        \centering
        \begin{tikzpicture}[scale=1.0, remember picture]
    \begin{axis}[
        ylabel=$\textrm{Power}$ (kW),
        ymin=0,ymax=50,
        xmin=0.8, xmax=5.5,
        xticklabels={,,12:00,,,,12:01,},  
        yticklabels={,,,,},
        legend style={nodes={scale=0.5}},
        domain=1:5
      ]
      \pgfmathsetseed{1}
      \addplot[name path=Baseline, smooth,black,thick, dashed] plot coordinates {
          (1,10)
          (2,12)
          (3,12.3)
          (4,13)
          (5,11.9)
        };
       \addlegendentry{Baseline}
       \addplot[name path=realized, smooth, red, mark=x, mark size=1.5pt] plot coordinates {
          (1,10)
          (1.5,11.5)
          (2,12)
          (2.4,10)
          (3,12.3)
          (3.6, 13)
          (4,13)
          (4.4,12.7)
          (5,11.0)
        };
      \addlegendentry{Realized}

      \node (source) at (axis cs:3.6,12.8) {};
      \node (destination) at (axis cs:2.9,30){};
      \draw[->](source)--(destination);

      \node[anchor=south] (source) at (axis cs:2.3,2.5) {\scriptsize Up-regulation};
      \node (destination) at (axis cs:2.4,9.5){};
      \draw[->](source)--(destination);
      
      \addplot[only marks, mark=x, orange, mark size=3.0pt] coordinates {(2,12) (3,12.3) (4,13)};
      \addlegendentry{50 Hz}

      \addplot[blue!50, opacity=0.2] fill between[of=Baseline and realized];
      \addlegendentry{Energy}
      
      \coordinate (insetPosition) at (rel axis cs:0.05,0.95);

    \end{axis}
    \begin{axis}[
      at={(insetPosition)},anchor={outer north west},footnotesize,
        ymin=-1,ymax=1,
        xmin=49.8, xmax=50.2,
        domain=49.8:50.2,
        xtick={49.8,50,50.2},
        xticklabels={49.8,{50 Hz},50.2},
        ytick={-1,0,1},
        yticklabels={\scalebox{0.7}{$-\textrm{P}^{\downarrow}$},\scalebox{0.7}{Baseline},\scalebox{0.7}{$+\textrm{P}^{\uparrow}$}},      
        legend style={nodes={scale=0.5}},
      ]
    \addplot[smooth, cyan, forget plot] {-5*x+250};
    \pgfmathsetseed{3}
    \addplot[only marks, mark=x, red, mark size=1.5pt] {(-5*x+250)+0.1*rand};
    \addlegendentry{Measurements};
    \addplot[name path=A, smooth, cyan, opacity=0.1,forget plot] {-5*(x-0.010)+250};
    \addplot[name path=B, smooth, cyan, opacity=0.1,forget plot] {-5*(x+0.010)+250};
    \addplot[cyan!50, opacity=0.9] fill between[of=A and B];
    \addlegendentry{Frequency band}
    \end{axis}
\end{tikzpicture}
        \caption{A one-minute snapshot of a demand portfolio reserved for FCR between hours 12:00 and 13:00. Since no schedule exists, a reference baseline is forecasted at 12:00 one hour into the future. A controller ensures that the realized consumption follows the reference at all times, effectively meaning that it crosses the reference whenever the frequency is 50 Hz. This allows for verification of the frequency response as shown in the inner figure.}
        \label{fig:FCR_demand_continuous_control_baseline_v2}
    \end{subfigure}
    \hfill
    \begin{subfigure}[t]{0.49\textwidth}
        \centering
        \begin{tikzpicture}[scale=1.0]
    \begin{axis}[
        ylabel=$\textrm{Power}$ (kW),
        ymin=0,ymax=20,
        xmin=0, xmax=6.5,
        xticklabels={,,12:00,,,,13:00,},
        yticklabels={,,,,},
        width = \axisdefaultwidth,
        domain=1:5     
     ]
    \addplot[smooth,red,thick] plot coordinates {
        (1,10)
        (1.2,9.2)
        (1.4,9)
        (1.6,8)
        (1.92,8.5)  
    };
    \addlegendentry{Realized}
    \addplot[name path=realized, smooth,red,thick, forget plot] plot coordinates {
        (1.92,8.5)
        (2,2.2)
        (3,2)
        (3.5,2)
        (3.7,8)
    };
    \addplot[smooth,red,thick, forget plot] plot coordinates {
        (3.7,8)
        (3.9,8.4)
        (4,8.2)
        (4.5,7.8)
        (5,9.1)  
    };
    \addplot[name path=Baseline, smooth, red, black, very thick, dashed] plot coordinates {
        (1.92,8.5)
        (2.5,8.6)
        (3,7.9)
        (3.5,7.6)
        (3.7,8)
    };
    \addlegendentry{Baseline}
    \addplot[blue, opacity=0.2] fill between[of=Baseline and realized];
    \addlegendentry{Energy}

    \end{axis}

    \draw [decorate,decoration={brace}]
    (1.8,0.5) -- (1.8,2.2) node [black,midway,xshift=-0.4cm] 
    {\footnotesize $\textrm{P}^{\uparrow}$};
 \end{tikzpicture}
        \caption{mFRR up-regulation between hours 12:00 and 13:00 for a flexible demand portfolio. The reserve capacity for demand is $\textrm{P}^{\uparrow}$, and a baseline forecast is made at the start of the activation period (dashed line). Delivery should be with respect to the baseline forecast made from the current consumption (red line), assuming non-continuous control.}
        \label{fig:mFRR_demand_upregulation_v2}
    \end{subfigure}
    \caption{A sample activation of a demand portfolio for FCR (left plot) and mFRR (right plot) services  between hours 12:00 and 13:00. Baseline: self-reported forecasted baseline. Realized: realized power output of the generator after reserve activation. Energy: energy delivered for reserve provision.}
    \label{fig:demand_flexibility}
\end{figure}

In the case of FCR as depicted in Figure \ref{fig:FCR_demand_continuous_control_baseline_v2}, a self-reported \textit{forecasted} baseline of the aggregate demand in the portfolio is used as the reference from which the frequency response is made. The forecast must be made in the beginning of the reservation hour, i.e., at 12:00 in Figure \ref{fig:FCR_demand_continuous_control_baseline_v2}, and the flexible capacity, $\textrm{P}^{\uparrow}$ and $\textrm{P}^{\downarrow}$, must be delivered with respect to the forecasted baseline. It can then be verified ex-post that the response adhered to the required response profile. This is shown in the inner figure where each time point is represented by a frequency in the grid and an aggregate power level which must lie in the acceptable frequency response band shown in blue. The forecasted baseline can be updated during the reservation hour as well, otherwise situations might arise where the flexible energy of the portfolio is used solely for following the forecasted baseline \cite{biegel2014integration}.

Similar to FCR, in the case of mFRR as illustrated in Figure \ref{fig:mFRR_demand_upregulation_v2}, the baseline should be \textit{forecasted} at the activation time. A verification can then be performed ex-post. Contrary to the case of FCR, only the average response in the activation period is considered, since the verification is done on the integral between the baseline and the response, i.e., the energy delivered. The baseline forecast can in principle be computed ex-post, but it should be computed ex-ante such that the aggregator ensures that the response adheres to the requirements.



\subsubsection{Key differences between FCR and mFRR provision}

The flexible capacity of a demand portfolio, $\textrm{P}^{\uparrow}$, must be delivered with a large degree of certainty no matter what the realization of the portfolio consumption is at the activation time. This applies both for FCR and mFRR provision. This stochasticity is crucial to model from the aggregator's point of view, and it is also important that the TSO can accommodate it. In Denmark, the TSO expects that a minimum of 90\% of all reserve capacity bids can be fully delivered \cite{energinet:prequalification}. It is still possible to utilize the day-ahead and intra-day schedule as the reference if it is known for aggregators as in Figure \ref{fig:generator_flexibility}.

The main difference between the baseline used for FCR and mFRR provision, is that the FCR provision requires a continuous control of the portfolio, and therefore must use the forecasted baseline as a 50 Hz baseline ex-ante due to the very particular response required for frequency deviations. Hence, it is also in the aggregator's own interest to use the best possible forecast, since most of the flexibility would otherwise be used to follow an inaccurate baseline \cite{biegel2014integration}. For mFRR, the forecast could potentially be done ex-post since the duration of a directional response is much longer and simpler. However, there might be an incentive to compute a biased baseline in this case. Other methods than forecasted baselines can be used as will be explained later.

\subsubsection{Incentives for flexibility provision in Market Model 3.0}

Market Model 3.0 ultimately encourages flexibility by lowering barriers for new actors to aggregate and provide their flexibility who otherwise would not. Hence, Market Model 3.0 has aimed to incentivize new market participants by lowering legal and business-related barriers to entry with respect to the provision of ancillary services. This applies not only for demand-side flexibility but also supply-side flexibility from intermittent renewable energy sources such as solar and wind power, where the same methodology applies.  

In the upcoming years, the European power grid will be unified with respect to the secondary and tertiary reserves with the formation of continental ancillary service markets, called the Platform for the International Coordination of Automated Frequency Restoration and Stable System Operation (PICASSO) \cite{PICASSO} and the Manually Activated Reserves Initiative (MARI) \cite{MARI}, as well as a common platform for replacement reserves called Trans European Replacement Reserves Exchange (TERRE) \cite{TERRE}. This will most likely increase the competition among flexibility service providers, since aggregators will have access to a much bigger and more liquid ancillary service markets than incumbent markets in each local synchronous area. It is expected to further incentivize new aggregators and current market participants to enter the ancillary service markets.

Demand-side flexibility needs to be profitable from consumers, aggregators, and the TSO's points of view. Economic incentives should also be considered when assessing barriers to demand-side flexibility \cite{biegel2014value}. Subsequently, it still remains a challenge to properly characterize demand-side flexibility, and many have tried creating a generic framework for flexibility provision \cite{maasoumy2014model, Harbo2013, biegel2013information}. Given a proper characterization of the flexibility, an economic assessment should be carried out, which serves as a basis for an incentive model on investment in flexibility provision.

\subsubsection{Additional benefits in Market Model 3.0}

There are several other initiatives in Market Model 3.0 that are focused on other issues than the role of aggregators. Local flexibility at the distributional level requires a new legislation that alters the revenue structure for Danish DSOs, who currently have a strong incentive to reinforce the grid as the only reaction to the increasing electrification. Furthermore, Market Model 3.0 emphasizes on data transparency especially on the low-voltage distribution grid. It is believed that the increased transparency will foster new business models and innovation. However, the DSO-TSO coordination is still a challenge. There is a focus on energy communities as well. Specifically, multi-meter and distributed communities that act as non-profit aggregators are suggested as the most viable option, and the legislation should facilitate and incentivize the formation of such a type of energy communities.

\section{Market Model 3.0: From concept to implementation}

While Market Model 3.0 mitigates the business and regulation complexity around the aggregator's role, it brings technical challenges from the aggregator's and the TSO's perspectives. However, the majority of those challenges are not new or unique, and have been addressed in many ongoing research activities.

\subsection{Baseline forecasting}

In the context of the Market Model 3.0, it cannot be assumed that the operating schedule of the demand is known in advance. Hence, as shown previously in Figures \ref{fig:FCR_demand_continuous_control_baseline_v2} and \ref{fig:mFRR_demand_upregulation_v2}, a new baseline reference is needed for the verification of delivery and subsequently payments. Demand forecasting at the activation time constitutes the baseline forecast, and the aggregator needs to do so accurately and document it to a satisfactory degree for the TSO when providing ancillary services. Baseline forecasts are known to be erroneous \cite{ziras2021baselines} and provide perverse incentives for aggregators to alter their forecast \cite{Borenstein2002, muthirayan2019mechanism}. Furthermore, it is likely not within the TSO's capability to control and verify each and every aggregator's own methodology for forecasting the baseline. They have to rely on a certain set of accuracy metrics provided by the aggregator to be assured that the forecast model is precise enough.

These problems mainly arise for mFRR, which requires a slow and often long unidirectional response. As mentioned for FCR, a forecasted baseline is not a problem in this regard, assuming a continuous control is exercised on the demand portfolio during the reservation hour. On the contrary, the aggregator has an incentive to forecast the portfolio as accurately as possible such that the use of demand flexibility is ensured for providing ancillary services, and not solely for tracking the baseline \cite{biegel2014integration}.

This topic should be studied further, including other potential mechanisms for generating the baseline reference. One potential mechanism is the one suggested by \cite{schneider2021online}, where an online learning mechanism is used for determining the power contribution. Another potential mechanism is introduced in \cite{Recurve} where an open-source tool works as an avoided cost calculator that quantifies the demand response using a reference group as a comparison.

Furthermore, there should be transparency and consistency in the response baseline estimation of a demand portfolio consisting of multiple individual assets. Imagine a response carried out by 50 assets in a total portfolio of 200 assets. Then the baseline should be estimated using exactly the same 50 assets. Otherwise, aggregators are prone to manipulate the baseline estimation to hide an insufficient response.

\subsection{TSO-BRP compensation mechanism}
As illustrated in Figure \ref{fig:aggregator_role_marketmodel}, Market Model 3.0 allows the aggregator to offer demand-side flexibility from consumers that each has a different BRP. Consequently, the aggregator may cause imbalances to the BRPs for ancillary services with non-negligible energy delivery. This incurred imbalance somehow needs to be accounted for in the balance settlement between BRPs and aggregators. A study was published on compensation mechanisms for aggregators \cite{NordicEnergyResearch} which looked at different high-level implementations in seven European countries with a focus on three different settlement models. However, it is not mentioned how the settlement models take into account potential overlapping between aggregators and BRPs as in Figure \ref{fig:aggregator_role_marketmodel}, since it is still assumed that there is no overlap. To the best of our knowledge, there is no such a compensation or a payment mechanism in the literature yet that has studied the intricacies of multiple large demand portfolios that incurs imbalances to multiple BRPs. Such a settlement scheme must be universal and transparent for the participants. Meanwhile, it requires that the aggregator is able to disaggregate the flexibility provision to assign contributions to each affected BRP. This will rely on an internal model of the aggregator that assigns individual contributions of each asset or a group of assets per BRP, e.g., through response tracking error \cite{bondy2016procedure} or other performance indices \cite{bondy2014performance}.

\subsection{Accounting for the rebound effect}

For thermostatically controlled loads, there is an inevitable rebound effect after an activation \cite{greening2000energy}, \cite{sorrell2008rebound}. It thus causes an imbalance in subsequent hours which the aggregator and/or BRP must be responsible for. This present two issues. First, the aforementioned compensation mechanism should consider the rebound. Second, the rebound reduces the potential of demand-side flexibility due to the imbalance after activation. It is therefore necessary to specify how a rebound should be handled between the aggregators, BRPs, and the TSO.

\subsection{Determination of reserve capacity}

Given how ancillary services are functioning, it is necessary for the aggregator to estimate the day before how much reserve capacity can be offered, e.g., for FCR or mFRR\footnote{For mFRR, it is also possible to participate only in the real-time balancing market.}. This is also a challenge for the TSO to somehow verify that the aggregator is not overestimating the capacity bids. Determining the reserve capacity the day before is a difficult problem for a (heterogeneous) demand portfolio, as it is NP-complete in nature \cite{petersen2014heuristic}, and requires a thorough and physical understanding of the underlying assets in the demand portfolio. From the TSO's point of view, it should ideally be straightforward and transparent to verify that aggregators determine and bid their true reserve capacity. This topic is therefore interesting to investigate further, both from an aggregator's and a TSO's perspective.


\subsection{Virtual metering}
There is big potential to utilize appliances with no sub-meters, as it increases the available demand-side flexible capacity. In Market Model 3.0, it is outlined how the legal requirement for sub-meters can be relaxed if it is proven that the power consumption can be estimated fairly accurately. In practice, this means the aggregator has to create power curves of a known set of assets with sub-meters (e.g., power curves of fan speed and consumption for an air handler unit) that can then be used for a set of similar assets with no sub-meters. The Danish TSO is currently conducting a pilot project about this possibility, called \textit{virtual metering} \cite{energinet_virtual_meter_pilot} of which, e.g., the company IBM is participating in. However, it eventually introduces more uncertainty to the delivery of ancillary services, as the demand-side portfolio consumption is suddenly not known fully from measurements, but instead partly estimated. How a heterogeneous demand-side portfolio behaves with a varying proportion of virtually-metered devices and how much uncertainty it introduces are both unknown. It is a further topic of research that seems largely untouched by the research community.

\subsection{Pre-qualification of reserves}

Given the above implications of Market Model 3.0, the official pre-qualification procedure of ancillary services must be changed accordingly. The pre-qualification is traditionally done with a specific focus on the flexibility provision by conventional generators \cite{bondy2018redefining}. However, it is now necessary to allow for stochastic demand-side flexibility as well \cite{bondy2017performance, bondy2016procedure}. This gives extra complexity in the verification and pre-qualification of uncertain portfolios, while a simple and transparent solution needs to be implemented as the pre-qualification itself should not be a hindrance for demand-side flexibility penetration.

\subsection{Additional incentives}

Demand response provides a clean and emission-less service to the power grid, as opposed to flexible, conventional, fossil-fueled based generators. This gives an additional incentive to become an aggregator with flexible demand. However, it is generally complex to directly quantify, and therefore monetize the expected savings in carbon-dioxide. It requires full knowledge of all flexibility providers bidding for reservation capacity and subsequent activation in each hour to determine the marginal service provider being displaced by flexible demand. For example, the benefit is large if a coal-based power plant is displaced by a portfolio of flexible demand when bidding for mFRR reserve capacity, but the benefit is zero if a hydro-based power plant is being displaced. Hence, an effort needs to be made to quantify, and even monetize, emission reductions from flexible demands.

\section{Conclusion}

This paper has described the challenges faced by the future power system with an increased electrification and decarbonization, and how demand-side flexibility will play an important role in this transition. To incentivize and increase penetration of demand-side flexibility, Denmark has recently released Market Model 3.0, which is a market framework that inherently integrates the aggregator as an independent participant in the power system. This will mitigate the barrier-of-entry for new aggregators and make it more attractive to utilize demand-side flexibility. 


In the vision of Market Model 3.0,  aggregators will be independent, such that the process for the provision of ancillary services will be changed with respect to pre-qualification, delivery, and verification. Currently, it is always assumed that a provider of e.g., FCR or mFRR, has a  reference schedule from which the response to ancillary services can be measured. In Market Model 3.0, this is not necessarily true anymore for aggregators. Hence, aggregators face additional technical challenges when utilizing demand-side flexibility such as determination of reserve capacity and pre-qualification. 

By mitigating market and legislation barriers for aggregators, Market Model 3.0 gives rise to other technical challenges such as skewed incentives for aggregators and the need for imbalance compensation. However, Market Model 3.0 is a pioneering effort by Denmark to address the green transition in the power system which is ambitious and groundbreaking, finally unlocking demand-side flexibility and the value it can bring to  society. 

\section*{Acknowledgement}

The authors would like to acknowledge the financial support from Innovation Fund Denmark under grant number 0153-00205B for partially funding the work in this paper. The authors would also like to thank Christian Ahn Albertsen (IBM Client Innovation Center) for all our discussions and his feedback which has been particularly valuable regarding practical challenges faced for demand-response. The authors would also like to thank Bennevis Crowley (DTU) for reading the paper and providing feedback on corrections, figures, and phrases. The authors would also like to thank the Centre for Utilities and Supply at Energistyrelsen (Danish Energy Agency) for allowing us to present our work to them while also elaborating their perspective on Market Model 3.0 and their intentions.


\printbibliography

@book{morales2013integrating,
  title     = {Integrating renewables in electricity markets: operational problems},
  author    = {Morales, J. M. and Conejo, Antonio J and Madsen, Henrik and Pinson, Pierre and Zugno, Marco},
  volume    = {205},
  year      = {2013},
  publisher = {Springer Science \& Business Media}
}

@article{California,
  title        = {California has a plan to pay efficiency providers to help prevent blackouts},
  url = {https://www.canarymedia.com/articles/energy-efficiency/california-has-a-plan-to-pay-efficiency-providers-to-help-prevent-blackouts},
  note         = {Accessed: 2021-12-01},
  year = {2021},
  journal = {Canarymedia}
}

@article{Recurve,
  title        = {Flex value},
  url = {https://flexvalue.recurve.com/},
  note         = {Accessed: 2021-12-01},
  year = {2021},
  journal = {Recurve}
}

@article{WorldEnergyOutlookDemandResponse,
  title        = {Demand response},
  url = {https://www.iea.org/reports/demand-response},
  journal = {World Energy Outlook},
  note         = {Accessed: 2022-02-10},
  year = {2022}
}

@article{energinet:Systemydelser,
  title        = {Introduktion til Systemydelser},
  url = {https://energinet.dk/-/media/8F2B403B3A724197A179FF6DFD9A926C.pdf},
  note         = {Accessed: 2021-09-06},
  journal = {Energinet},
  year = {2021}
}

@article{energinet:GennemgangAfMarkeder,
  title        = {Gennemgang af Nuværende Systemydelsesmarkeder},
  url = {https://www.danskfjernvarme.dk/-/media/danskfjernvarme/kurser_og_arrangementer/modematerialer/temamoder/2021/2021-06-02-udnyt-elmarkederne-og-tjen-penge/gennemgang-af-markeder_kathrine-lau-j%C3%B8rgensen,-energinet.pdf},
  note         = {Accessed: 2022-08-04},
  journal = {Energinet},
  year = {2022}
}

@article{NordicBalancingModel,
  title        = {The Nordic Balancing Model},
  url = {https://nordicbalancingmodel.net/roadmap-and-projects/},
  note         = {Accessed: 2021-11-23},
    year = {2021},
    journal = {nordicbalancingmodel.net}
}

@article{NordicEnergyResearch,
  title        = {The regulation of independent aggregators with a focus on compensation mechanisms},
  url = {https://www.nordicenergy.org/article/the-regulation-of-independent-aggregators/},
  note         = {Accessed: 2022-08-02},
    year = {2022},
    journal = {nordicenergy.org}
}

@article{MARI,
  title        = {MARI},
  url = {https://www.entsoe.eu/network_codes/eb/mari/},
  note         = {Accessed: 2021-11-23},
  year = {2021},
  journal = {ENTSOE}
}

@article{PICASSO,
  title        = {PICASSO},
  url = {https://www.entsoe.eu/network_codes/eb/picasso/},
  note         = {Accessed: 2021-11-23},
  year = {2021},
    journal = {ENTSOE}
}

@article{TERRE,
  title        = {PICASSO},
  url = {https://www.entsoe.eu/network_codes/eb/terre/},
  note         = {Accessed: 2022-08-17},
  year = {2022},
    journal = {ENTSOE}
}

@article{TENNETNL,
  title        = {TenneT Flexibility Monitor},
  url = {https://www.tennet.eu/fileadmin/user_upload/Company/Publications/Technical_Publications/Dutch/20200117_TenneT_Flexibility_Monitor.pdf},
  note         = {Accessed: 2021-11-29},
  year = {2021},
  journal = {TenneT}
}

@article{energinet:BRP_list,
  title        = {Balance responsible parties in Denmark},
  url = {https://en.energinet.dk/Electricity/New-player/Oversigt-over-BA},
  note         = {Accessed: 2021-10-15},
  year = {2021},
  journal = {Energinet}
}

@article{energinet:prequalification,
  title        = {Prequalification of units and aggregated portfolios},
  url = {https://en.energinet.dk/Electricity/Ancillary-Services/Prequalification-and-test},
  note         = {Accessed 2021-09-22},
  year = {2021},
  journal = {Energinet}
}

@article{energistyrelse:Marketmodel3.0,
  title        = {Market Model 3.0},
  url = {https://ens.dk/en/our-responsibilities/electricity/market-model-30},
  note         = {Accessed 2021-10-15},
  year = {2021},
  journal = {Danish Energy Agency}
}

@article{energinet_virtual_meter_pilot,
  title        = {Pilotprojekt til test af virtuelle målere til validering af en systemydelsesrespons},
  url = {https://energinet.dk/El/Systemydelser/Nyheder-om-systemydelser/2021-03-24-Virtuel-maeling-pilotprojekt},
  note         = {2021-10-08},
  year = {2021},
  journal = {Energinet}
}

@article{Ostergaard2021,
  author    = {J. Østergaard and Charalampos Ziras and Henrik W. Bindner and Jalal Kazempour and Mattia Marinelli and Peter Markussen and Signe Horn Rosted and Jorgen S. Christensen},
  journal   = {IEEE Power and Energy Magazine},
  pages     = {46-55},
  title     = {Energy security through demand-side flexibility: The case of Denmark},
  number    = {2},
  volume    = {19},
  year      = {2021},
}

@article{Harbo2013,
  author   = {S. Harbo and others},
  journal  = {IEEE/PES Innovative Smart Grid Technologies Europe, ISGT-Europe 2013},
  pages = {1-5},
  title    = {Contracting flexibility services},
  year     = {2013}
}

@article{Borenstein2002,
  author = {S. Borenstein and M. Jaske and A. Rosenfeld and E. Org},
  %note   = {Really important discussion of baselines.},
  title  = {Dynamic pricing, advanced Metering, and demand response in electricity Markets},
  year   = {2002}
}

@inproceedings{NordicBalancingMarkets,
  author    = {Khodadadi, A. and Herre, Lars and Shinde, Priyanka and Eriksson, Robert and Söder, Lennart and Amelin, Mikael},
  booktitle = {2020 17th International Conference on the European Energy Market (EEM)},
  title     = {Nordic Balancing Markets: Overview of Market Rules},
  year      = {2020},
  volume    = {},
  number    = {},
  pages     = {1-6},
  %doi       = {10.1109/EEM49802.2020.9221992}
}

@article{ziras2021baselines,
  title     = {Why baselines are not suited for local flexibility markets?},
  author    = {Ziras, C. and others},
  journal   = {Renewable and Sustainable Energy Reviews},
  volume    = {135},
  note     = {Article No. 110357},
  year      = {2021}
}

@article{jin2020local,
  title     = {Local flexibility markets: Literature review on concepts, models and clearing methods},
  author    = {Jin, X. and others},
  journal   = {Applied Energy},
  volume    = {261},
  note     = {Article No. 114387},
  year      = {2020}
}

@inproceedings{bouloumpasis2019congestion,
  title        = {Congestion management using local flexibility markets: Recent development and challenges},
  author       = {Bouloumpasis, I. and others},
  booktitle    = {2019 IEEE PES Innovative Smart Grid Technologies Europe (ISGT-Europe)},
  pages        = {1--5},
  year         = {2019}
}

@article{kirby2007ancillary,
  title   = {Ancillary services: Technical and commercial insights},
  author  = {Kirby, B.},
  journal = {Retrieved October},
  volume  = {4},
  pages   = {2012},
  year    = {2007}
}

@article{biegel2014value,
  title     = {Value of flexible consumption in the electricity markets},
  author    = {Biegel, B. and Hansen, Lars Henrik and Stoustrup, Jakob and Andersen, Palle and Harbo, Silas},
  journal   = {Energy},
  volume    = {66},
  pages     = {354--362},
  year      = {2014},
  publisher = {Elsevier}
}

@article{muthirayan2019mechanism,
  title     = {Mechanism design for demand response programs},
  author    = {Muthirayan, D. and Kalathil, D. and Poolla, K. and Varaiya, P.},
  journal   = {IEEE Transactions on Smart Grid},
  volume    = {11},
  number    = {1},
  pages     = {61--73},
  year      = {2019}
}

@article{petersen2014heuristic,
  title     = {Heuristic optimization for the discrete virtual power plant dispatch problem},
  author    = {Petersen, M. K. and Hansen, Lars H and Bendtsen, Jan and Edlund, Kristian and Stoustrup, Jakob},
  journal   = {IEEE Transactions on Smart Grid},
  volume    = {5},
  number    = {6},
  pages     = {2910--2918},
  year      = {2014},
  publisher = {IEEE}
}

@inproceedings{maasoumy2014model,
  title        = {Model predictive control approach to online computation of demand-side flexibility of commercial buildings hvac systems for supply following},
  author       = {Maasoumy, M. and Rosenberg, C. and Sangiovanni-Vincentelli, A. and Callaway, D. S.},
  booktitle    = {2014 American Control Conference},
  pages        = {1082--1089},
  year         = {2014}
}

@inproceedings{biegel2013information,
  title        = {Information modeling for direct control of distributed energy resources},
  author       = {Biegel, B. and Andersen, Palle and Stoustrup, Jakob and Hansen, Lars Henrik and Tackie, David Victor},
  booktitle    = {American Control Conference},
  pages        = {3498--3504},
  year         = {2013}
}

@article{schneider2021online,
  title     = {An online learning framework for targeting demand response customers},
  author    = {Schneider, I. and others},
  journal   = {IEEE Transactions on Smart Grid},
  year      = {2022},
  volume = {13},
  number = {1},
  pages = {293-301}
}

@inproceedings{bondy2018redefining,
  title     = {Redefining Requirements of Ancillary Services for Technology Agnostic Sources},
  author    = {Morales Bondy, D. E. and MacDonald, J. and Kara, E. C. and Gehrke, O. and Heussen, K. and Chassin, D. and Kiliccote, S. and Bindner, H. W.},
  booktitle = {Proceedings of the 51st Hawaii International Conference on System Sciences},
  pages = {1-10},
  year      = {2018}
}

@inproceedings{bondy2017performance,
  title        = {Performance requirements modeling and assessment for active power ancillary services},
  author       = {Morales Bondy, D. E. and Thavlov, A. and Tougaard, Janus Bundsgaard Mosb{\ae}k and Heussen, Kai},
  booktitle    = {IEEE Manchester PowerTech Conference},
  pages        = {1--6},
  year         = {2017}
}

@inproceedings{bondy2016procedure,
  title        = {Procedure for validation of aggregators providing demand response},
  author       = {Morales Bondy, D. Esteban  and Gehrke, O. and Thavlov, A. and Heussen, K. and Kosek, Anna M and Bindner, Henrik W},
  booktitle    = {Power Systems Computation Conference (PSCC)},
  pages        = {1--7},
  year         = {2016}
}

@inproceedings{bondy2014performance,
  title        = {Performance assessment of aggregation control services for demand response},
  author       = {Morales Bondy, D. E.  and Costanzo, G. T. and Heussen, K. and Bindner, H. W.},
  booktitle    = {IEEE PES Innovative Smart Grid Technologies, Europe},
  pages        = {1--6},
  year         = {2014}
}

@article{biegel2014integration,
  title     = {Integration of flexible consumers in the ancillary service markets},
  author    = {Biegel, B. and Westenholz, M. and Hansen, L. H. and Stoustrup, J. and Andersen, P. and Harbo, S.},
  journal   = {Energy},
  volume    = {67},
  pages     = {479--489},
  year      = {2014}
}

@inproceedings{petersen2013market,
  title        = {Market integration of virtual power plants},
  author       = {Petersen, M. K. and Hansen, Lars Henrik and Bendtsen, J and Edlund, Kristian and Stoustrup, Jakob},
  booktitle    = {52nd IEEE Conference on Decision and Control},
  pages        = {2319--2325},
  year         = {2013}
}

@article{sorrell2008rebound,
  title={The rebound effect: Microeconomic definitions, limitations and extensions},
  author={Sorrell, S. and Dimitropoulos, J.},
  journal={Ecological Economics},
  volume={65},
  pages={636--649},
  year={2008},
  publisher={Elsevier}
}

@article{greening2000energy,
  title={Energy efficiency and consumption—The rebound effect—A survey},
  author={Greening, L. A. and Greene, D. L. and Difiglio, C.},
  journal={Energy policy},
  volume={28},
  pages={389--401},
  year={2000},
  publisher={Elsevier}
}


\section*{Author Biographies}

\textbf{Peter A.V. Gade} is an Industrial PhD researcher at IBM and affiliated with the Technical University of Denmark, Kongens Lyngby, Denmark, in the Energy Markets and Analytics Section within the Power and Energy Systems division at the Wind and Energy Systems Department. His research focuses on demand-side flexibility and the revenue streams from utilization of demand-side flexibility. He holds a M.S. in Mathematical Modelling and Computing and a B.S. in Biomedical Engineering, both from the Technical University of Denmark.
\\
\\
\textbf{Trygve Skjøtskfit} is an Associate Partner at IBM Denmark, with focus on energy transformation and demand-side flexibility. His solid experience and deep knowledge within intelligent energy systems, buildings, and civil infrastructures makes him a leading figure, strategic advisor, and a first mover in the flexibility market with a strong track record to find and deliver new cutting-edge solutions. He holds an MBA in Strategy from Universitat Pompeu Fabra, and a Master of Export Engineering from Copenhagen University, College of Engineering.
\\
\\
\textbf{Henrik W. Bindner} received the MSc in Electrical Engineering from Technical University of Denmark in 1988. He is currently a senior researcher with the Department of Wind and Energy Systems, Technical University of Denmark. He is heading the \textit{Distributed Energy Systems} Section and his research interests include control and management of smart grids, active distribution networks, and integrated energy systems.
\\
\\
\textbf{Jalal Kazempour}  is an Associate Professor with the Department of Wind and Energy Systems, Technical University of Denmark, where he is heading the \textit{Energy Markets and Analytics} Section. He received the Ph.D. degree in Electrical Engineering from the University of Castilla-La Mancha, Ciudad Real, Spain, in 2013. His research interests include intersection of multiple fields, including power and energy systems, electricity markets, optimization, game theory, and machine learning.

\end{document}